\documentclass[preprint,proceedings]{rmaa}

\SetYear{2004} \SetConfTitle{RevMexAA (Serie de Conferencias), 21,
15-19}

\title{Spin Correlation In Binary Systems}

\author{N. Farbiash \altaffilmark{1} and R. Steinitz \altaffilmark{1}}

\altaffiltext{1}{Ben-Gurion University, Beer-Sheva, Israel 84105}

\shortauthor{Farbiash \& Steinitz}

\shorttitle{Spin Correlation in Binary Systems}

\fulladdresses
{
\item Ben-Gurion University, Beer-Sheva, Israel
84105 (\email{farbiash@bgumail.bgu.ac.il ,
steinitz@bgumail.bgu.ac.il})
}

\abstract{We examine the correlation of projected rotational
velocities in binary systems. It is an extension of previous work
(Steinitz and Pyper, 1970; Levato, 1974). An enlarged data basis
and new tests enable us to conclude that there is indeed
correlation between the projected rotational velocities of
components of binaries. In fact we suggest that spins are already
correlated.}

\addkeyword{Stars: Binaries-General} \addkeyword{Stars:
Rotational-velocities-distribution} \addkeyword{Stars: Spin
Correlation}

\begin{document}

\maketitle

\section{Introduction}
\label{sec:intro}

Evolution of binary systems could be a result of two main
processes: three body collision, or evolving of binaries in one
disk. The probability for three body collision is extremely small;
thus it is more likely that binary systems evolve in one disk. In
that case we expect to find correlation between the measured
$V_{e}sin(i)$ values of the members of such a system. Therefore,
we study the degree of projected rotational velocity correlation
between members of binary systems.

Slettebak (1963) did not find any significant difference between
mean rotational velocity distribution of members of binary systems
and those of single stars. Further, Abt (2001) concluded that spin
axes are probably randomly oriented. On the other hand, Steinitz
and Pyper (1970) concluded that some correlation of projected
rotational velocities is present for the components of visual
binaries. Levato (1974) also discussed this issue for visual
binaries and close binary systems and found that there is indeed
correlation of projected rotational velocities in binary systems.
We now extend the original study (Steinitz and Pyper, 1970), which
included only 50 systems. The significance of our results stems
from the use of 1010 binary systems.

Actually, we will examine three samples of binary systems, as
defined in the next section.

\section{Data}
\label{sec:data}

The Catalogue of Stellar Projected Rotational Velocities (Glebocki
et al, 2000) is our source for the sample of binaries, chosen by
imposing the following criteria:
\begin{enumerate}
    \item The spectral type of both components is earlier
than A9 (slow rotation of stars later than A9 would automatically
simulate correlation).
    \item Giants and Supergiants may have lost their original
rotational velocities. So we select only those binaries whose both
components are on the main sequence.
    \item Multiple systems including more than two stars are
excluded.
\end{enumerate}

We are now left with a sample of 1010 real binary (RB) systems.
Since the mean rotational velocity along the main sequence
changes, it could happen that the choice of a sample restricted in
spectral type will automatically exhibit the correlation we are
looking for. To eliminate this possibility, and show that
projected rotational velocity correlation is present in real
binary systems only, we use two artificial samples of binaries.
Define sample AB (Artificial Binaries) through shuffling the
components of the real systems, and exclude the real ones
(containing 2038180 systems). The second sample ABR (Artificial
Binaries, Restricted), is obtained from sample AB by eliminating
of all pairs having spectral type difference larger than two
spectral subclasses (containing 263344 systems). This restriction
is more stringent than the one we admit for the basic, real
systems.

\section{Illustration}
\label{sec:Illustration} In Fig.1. we plot the projected
rotational velocities of one component against the other one, for
all the three samples previously defined. From Fig.1. we see that
samples AB and ABR do not exhibit any correlation whatsoever. In
contrast, the correlation in sample RB is clearly evident. To
quantify this result we apply more tests in the next section.

\begin{figure}[!b]\centering
\includegraphics[width=0.99\columnwidth]{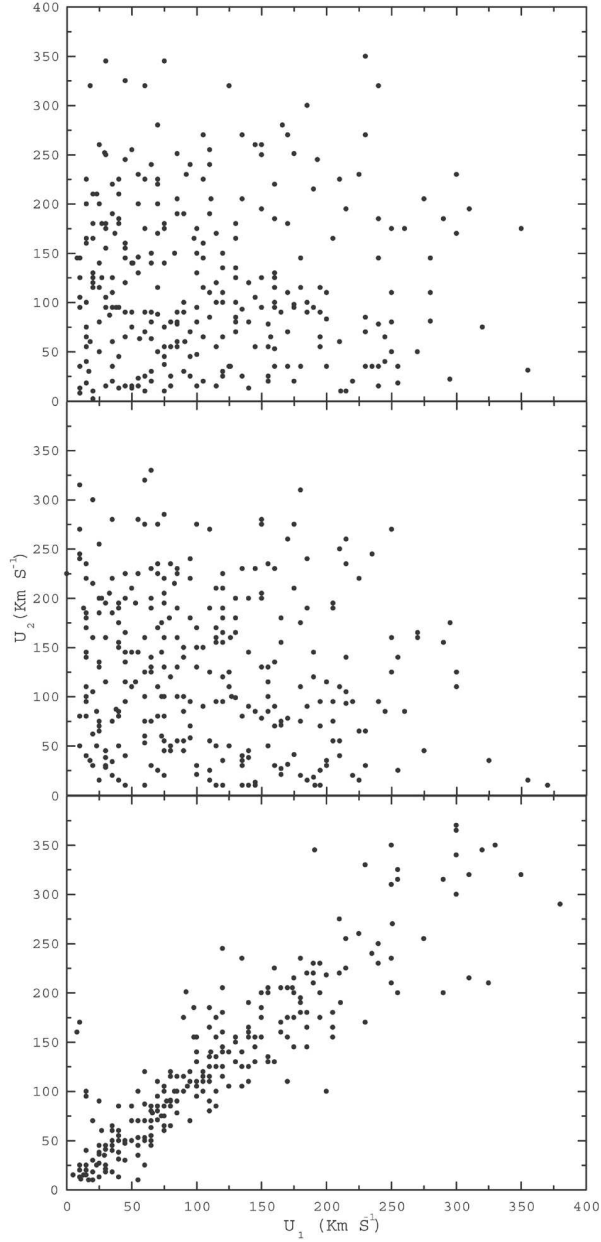}
  \caption{illustration of the projected
rotational velocities of one component against the other for the
samples AB (top), ABR (mid), and RB (bottom).}
 \label{fig1}
\end{figure}

\section{Analysis}
\label{sec:Analysis}

\subsection{Bivariate distribution, Marginal distribution, and Linear regression}

We prepare a table for each of the three samples, AB, ABR and RB:
it gives the discrete bivariate distribution of the samples. We
give also the regression of the mean velocity of one component on
the other one, as defined by equation (1).
\begin{equation}
\overline u _1 \left( {u_2 \left( j \right)} \right) =
\frac{\sum\limits_{i = 1}^7 {u_1 \left( i \right)F\left( {u_1
\left( i \right),u_2 \left( j \right)} \right)} }{\sum\limits_{i =
1}^7 {F\left( {u_1 \left( i \right),u_2 \left( j \right)} \right)}
} \
\end{equation}

Here $u_1=v_1sini_1$ is the projected rotational velocity of the
component, and $F(u_1 ,u_2 )$ is the bivariate distribution.

For better illustration we show the regression of projected
rotational velocity distribution of one component on the other one
for each of the three samples. The $F(u_1 ,u_2 )$ distributions
for the samples AB, ABR, and RB are given in table 1. The marginal
distribution and the regression lines are also shown. The latter
are plotted in Fig.2. For the ease of comparing the current
results with previous ones, the velocity range has been divided
into subintervals of 50 km s$^{- 1}$.

\begin{table*}[!t]\centering
  \newcommand{\DS}{\hspace{0.4\tabcolsep}} 
  \setlength{\tabnotewidth}{0.5\textwidth}
  \setlength{\tabcolsep}{0.9\tabcolsep}
  \tablecols{10}
  \caption{Discrete distribution function
 ${F(y_1,y_2)}$
 of the
 projected rotational velocities of the components of sample AB (top), ABR(middle), and RB(bottom).} \label{tab:AB}
  \begin{tabular}{l @{\DS} cccccccccc}
    \toprule
    $\begin{array}{cc}
   & u1 \\
   u2 & \\
\end{array}$ & 0-50 & 50-100 & 100-150 & 150-200 & 200-250 & 250-300 & 300-350 & $\phi
_2 (u_2 )$&
$\bar {u}_1 (u_2)$ \\
\midrule
0-50& 563& 525& 406& 351& 200& 90& 51& 0.22&
115 \\
50-100& 617& 575& 446& 388& 221& 99& 58& 0.24&
116 \\
100-150& 451& 420& 324& 282& 161& 73& 42& 0.18&
115 \\
150-200& 433& 404& 313& 271& 155& 70& 41& 0.17&
116 \\
200-250& 276& 260& 200& 174& 99& 45& 26& 0.11&
116 \\
250-300& 122& 115& 88& 77& 44& 20& 12& 0.05&
116 \\
300-350& 104& 99& 76& 67& 39& 17& 10& 0.04&
117 \\
$\phi _1 (u_1 )$& 0.26& 0.24& 0.19& 0.16& 0.09& 0.04& 0.02& &
 \\
$\bar {u}_2 (u_1)$& 125& 126& 126& 126& 126& 126& 127& &
 \\
 \midrule
0-50 & 576 & 496 & 385 & 356 & 205 & 105 & 0 & 0.21 &
112\\
50-100& 657& 627& 466& 412& 215& 94& 0& 0.25&
108 \\
100-150& 452& 447& 331& 277& 141& 56& 22& 0.17&
110 \\
150-200& 465& 454& 335& 276& 143& 57& 22& 0.18&
109 \\
200-250& 290& 264& 201& 172& 96& 45& 18& 0.11&
112 \\
250-300& 115& 99& 80& 65& 37& 19& 7& 0.04&
112 \\
300-350& 115& 88& 74& 66& 43& 25& 10& 0.04&
119 \\
$\phi _1 (u_1 )$& 0.27& 0.25& 0.19& 0.16& 0.09& 0.04& 0.01& &
 \\
$\bar {u}_2 (u_1 )$& 126& 125& 126& 124& 125& 125& 201& &
 \\
\midrule 0-50& 1901& 198& 20& 0& 0& 0& 0& 0.21&
31 \\
50-100& 614& 1564& 158& 40& 10& 0& 0& 0.24&
68 \\
100-150& 40& 515& 1000& 158& 0& 0& 0& 0.17&
112 \\
150-200& 69& 79& 574& 752& 158& 59& 10& 0.17&
156 \\
200-250& 10& 50& 99& 535& 347& 50& 40& 0.11&
190 \\
250-300& 0& 0& 30& 59& 228& 119& 30& 0.05&
231 \\
300-350& 0& 10& 10& 50& 129& 198& 89& 0.05&
254 \\
$\phi _1 (u_1 )$& 0.26& 0.24& 0.19& 0.16& 0.09& 0.04& 0.02& &
 \\
$\bar {u}_2 (u_1 )$& 43& 89& 144& 193& 242& 278& 284& &
 \\
      \bottomrule
    \tabnotetext{a}{$u_1$ and $u_2$ are given in units of $10^{-4}$ km s$^{- 1}$.}
  \end{tabular}
\end{table*}

\begin{figure}[!b]\centering
\includegraphics[width=0.99\columnwidth]{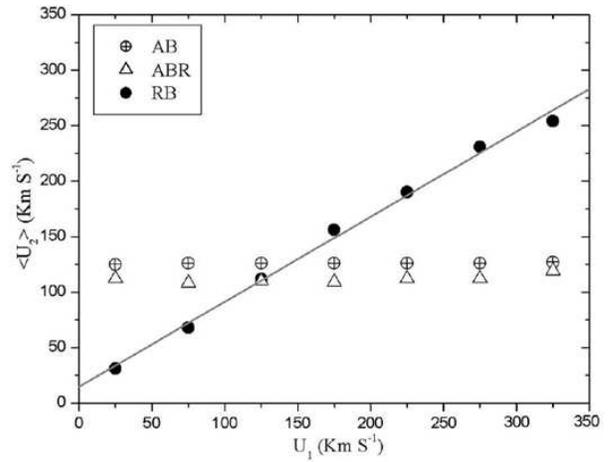}
  \caption{Regression of the mean rotational velocities of one component (ordinate) as a
  function of the rotational velocity interval of the second component (abscissa) for samples
   AB (top), ABR (mid), and RB (bottom).}
 \label{fig2}
\end{figure}

No significant differences between the main diagonal and other
components of the table are evident for the samples AB and ABR.
But the set RB shows a distinct difference between the main
diagonal and other elements of the table, indicating the presence
of correlation of projected spins in real binaries.

\subsection{Modified Convolution Test}

\begin{figure}[!ht]\centering
\includegraphics[width=0.99\columnwidth]{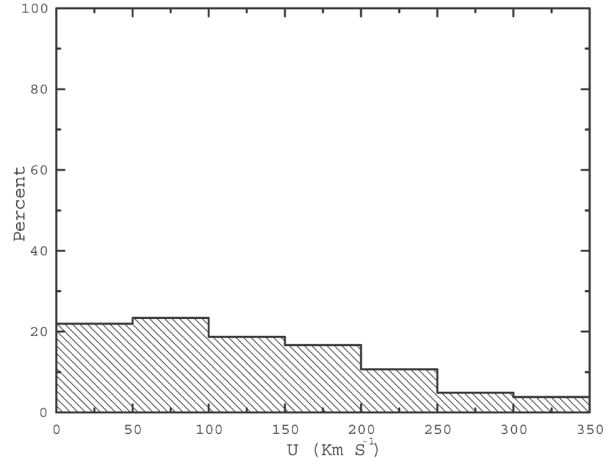}
 \caption{Projected rotational velocity distribution of all components grouped into intervals of 50 km s$^{- 1}$.}
 \label{fig3}
\end{figure}

\begin{figure}[!ht]\centering
\includegraphics[width=0.99\columnwidth]{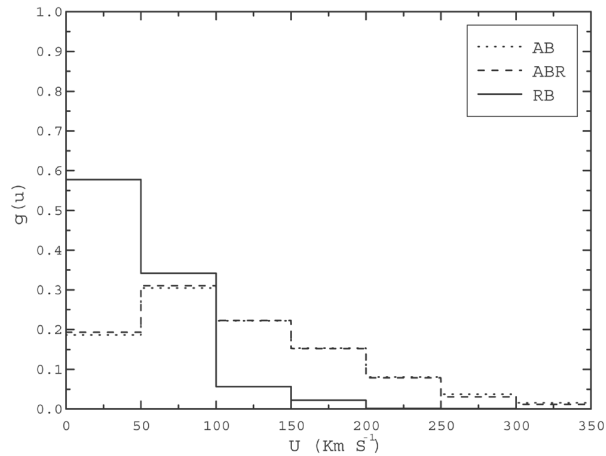}
 \caption{Modified convolution of rotational velocities.}
 \label{fig4}
\end{figure}

Further illustration of the correlation between the rotational
velocity distribution in real binary systems, and not in
artificial ones, is to evaluate the modified convolution of the
distribution functions. To validate the result, we first look at
the single star velocity distribution, as shown in Fig. 3. The
distribution is rather flat and does not indicate a strong maximum
at any specific speed. Our modified convolution is essentially {
$g(u) = \int{P(v - u|v)dv}$}, which in the discrete case is:

\begin{equation}
g(u) = \sum\limits_v {F(v - u,v)}
\end{equation}

As expected from the flatness of the distribution in Fig. 3., the
convolution of artificial binary systems do not indicate a
concentrated, sharp peak. However, for real binaries, there is a
rather sharp peak present. It means that probability for the
projected rotational velocity difference to be smaller than 50 km
s$^{- 1}$ is almost 60{\%}. The probability for the difference to
be smaller than 100 km s$^{- 1}$ is more than 90{\%}(!) in real
binaries.

\section{Conclusions}
\label{sec:Conclusions}

We examined three samples of binaries. For each of the samples we
applied identical tests, to render significant comparisons between
artificial binaries and real ones.

As expected, correlation in artificial systems (Set AB and ABR) is
insignificant. This is indicated by the miniscule slope of the
regression lines. However, correlation for the Binary set RB (Real
Binaries) is clearly evident ({u}$_{1}$={.856u}$_{2}$+32). One can
describe our results roughly as follows:

\begin{equation}
v_{1}\ sini_{1} \cong v_{2} \ sini_{2} \
\end{equation}

This relation can be understood either as:

\begin{enumerate}
    \item $v_{1} \ll\ v_{2}$ while $sini_{1} \gg\
sini_{2}$,

or
    \item $v_{1} \simeq\ v_{2}$ as well as $sini_{1}
\simeq\ sini_{2}$.
\end{enumerate}

Since we have used a sample containing 1010 systems, the
probability of the first case to be real is extremely small. We
rather accept the second explanation. We interpret this as meaning
that:

\begin{enumerate}
    \item Spin axes of members in binary systems are roughly parallel.
    \item Rotational speeds are correlated.
\end{enumerate}

There have been several investigations related to binary systems
(Giuricin et al., 1984; Levato, 1976; Zhan, 1977; and Pan, 1997).
These papers points out the importance of tidal interaction in
close binary systems (especially the theory given by Zhan, 1966,
1970, 1975, 1977).

We suppose that only a small fraction of the sample RB contains
close binary systems. Thus, we lack a general theory which can
account for the empirical results demonstrated here.

\end{document}